\documentclass[12pt]{article}
\usepackage{amsmath}
\usepackage{amssymb}
\usepackage{graphicx}
\usepackage{subfig}
\usepackage{float}
\begin{document}

\bigskip \ 

\bigskip \ 

\begin{center}
\textbf{Beyond Schwarzschild: New Pulsating Coordinates}

\smallskip

\textbf{for Spherically Symmetric Metrics}

\bigskip \ 

E. A. Le\'{o}n\footnote{%
ealeon@uas.edu.mx, corresponding author}, J. A. Nieto\footnote{%
niet@uas.edu.mx, janieto1@asu.edu}, A. Sandoval-Rodr\'{\i}guez\footnote{%
andres.fcfm@uas.edu.mx} and B. Mart\'{\i}nez-Olivas\footnote{%
brandon.fcfm@uas.edu.mx}

\bigskip \ 

\textit{Facultad de Ciencias F\'{\i}sico-Matem\'{a}ticas de la Universidad
Aut\'{o}noma}

\textit{de Sinaloa, 80010, Culiac\'{a}n, Sinaloa, M\'{e}xico.}

\bigskip \ 

\bigskip \ 

\textbf{Abstract}
\end{center}

\noindent Starting from a general transformation for spherically symmetric
metrics where g\_11=-1/g\_00, we analyze coordinates with the common
property of conformal flatness at constant solid angle element. Three
general possibilities arise: one where tortoise coordinate appears as the
unique solution, other that includes Kruskal-Szekeres coordinates as a very
specific case, but that also allows other similar transformations, and
finally a new set of coordinates with very different properties than the
other two. In particular, this represents any causal patch of the
spherically symmetric metrics in a compactified form. We analyze some
relations, taking the Schwarzschild case as prototype, but also contrasting
the cosmological de-Sitter and Anti-de-Sitter solutions for the new proposed
\textquotedblleft pulsating coordinates\textquotedblright .

\bigskip


Pacs numbers: 04.20.Cv, 04.20.Jb, 04.70.Bw, 98.80.-k

February 2024

\newpage

\noindent \textbf{1. Introduction}

\smallskip \ 

\noindent As it is well known, Schwarzschild spacetime can be described by
means of several coordinates. Some of those are more suitable for certain
type of observers, as is the case for Painleve-Gullstrand, others make
easier some connection with some limiting case, such as isotropic or the
very same Schwarzschild coordinates \cite{Dray}, while others allows to
obtain a visualization of the entire manifold, such as in Kruskal-Szekeres
coordinates or a Penrose-Carter compactification \cite{Kruskal}-\cite%
{Carroll}.

In this article, we analyze the case of a general transformation valid for
all spherically symmetric metrics where $g_{11}=-1/g_{00}$ \cite{Jacobson}%
\cite{Edgar}. We obtain two main interesting results: an appreciation of
tortoise and Kruskal-Szekeres coordinates as two limiting cases of more
general transformations, and the proposal and properties of a new coordinate
system, which we call \textit{pulsating coordinates}.

The structure of this article is the following. In Section 2, we review the
proposal that transforms the mentioned class of metrics to a conformal flat
(1+1)-form at constant angles, and the way that this implies three general
possibilities that give raise to several coordinate systems. In Section 3 we
review the simplest possibility, that yields directly the Regge-Wheeler, or
tortoise coordinates. We take a moment to digress about some properties of
these coordinates, as they are useful in the rest of the article. Section 4
deals with other natural solution that has a subclass of possible solutions.
We take the Schwarzschild case as prototype, and although Kruskal-Szekeres
coordinates are a justified selection, we show another case where interior
and exterior regions are well defined. We devote Section 5 to the last of
the general possible cases, as it implies a new coordinate system that has
some unexpected properties as compared with the previous ones, such as a
natural compactification and a type of pulsation of the surfaces that define 
$r$ and $t$ constant. There we mention as a case of interest the application
to Anti-de-Sitter solution, where the mentioned pulsation is unique.
Finally, in Section 6 we make some final remarks about the significance of
our approach, that differs with the traditional derivations of alternative
coordinate systems.\bigskip

\newpage

\noindent \textbf{2. (1+1)-conformal flat metrics for spherically symmetric
metrics with constant angles}.

\smallskip \ 

\noindent As it is well known, Einstein equations,%
\begin{equation}
R_{\mu \nu }-\frac{1}{2}g_{\mu \nu }(R-2\Lambda )=8\pi GT_{\mu \nu }, 
\tag{1}
\end{equation}%
have several analytic solutions that can be put in the form%
\begin{equation}
dS_{(1)}^{2}=-fdt^{2}+f^{-1}dr^{2}+r^{2}d\theta ^{2}+r^{2}\sin ^{2}\theta
d\phi ^{2},  \tag{2}
\end{equation}%
where $f=f(r)$. This encompass a broad class of metrics that have the
property of spherical symmetry, and that can be put in a static form.
Interestingly enough, some of this are a subclass of FLRW-metrics, such as
de Sitter (dS) and AdS space, while others describe the presence of black
holes, such as Schwarzschild, Reissner-Nordstr\"{o}m and
Schwarzschild-de-Sitter, just to name a few (see \cite{Edgar} and references
therein for details).

Consider a transformation of the metric to the form%
\begin{equation}
dS_{(2)}^{2}=\Omega (-dT^{2}+dX^{2})+r^{2}d\theta ^{2}+r^{2}\sin ^{2}\theta
d\phi ^{2},  \tag{3}
\end{equation}%
that is, we have a transformation to a conformal flat form in (1+1), when $%
d\theta =d\phi =0$. The existence of this transformation is guaranteed, and
this is properly justified in a simple way in the Appendix. Now consider the
transformation from $dS_{(1)}^{2}=g_{\alpha \beta }dx^{\alpha }dx^{\beta }$
given in (2) with coordinates $x^{\alpha }=(t,r,\theta \, \phi )$, to the
static spherical symmetric form $dS_{(2)}^{2}=\gamma _{\alpha \beta
}dy^{\alpha }dy^{\beta }$ given in (3), with $y^{\alpha }=(T,X,\theta ,\phi
) $. The angular part is the same, and then we shall be concerned with the
transformation

\begin{equation}
g_{ab}=\frac{\partial y^{a^{\prime }}}{\partial x^{a}}\frac{\partial
y^{b^{\prime }}}{\partial x^{b}}\gamma _{a^{\prime }b^{\prime }},  \tag{4}
\end{equation}%
where indices $a$ and $b$ run from $0$ to $1$. As $\Omega $, $T$ and $X$ are
functions of $t$ and $r$, the $g_{00}$, $g_{11}$ and $g_{01}$ components of
this equation are, after rearrangement, equivalent to%
\begin{equation}
\left( \partial _{t}X\right) ^{2}=\left( \partial _{t}T\right) ^{2}-f\Omega
^{-1},  \tag{5}
\end{equation}%
\begin{equation}
\left( \partial _{r}X\right) ^{2}=\left( \partial _{r}T\right)
^{2}+f^{-1}\Omega ^{-1}  \tag{6}
\end{equation}%
and%
\begin{equation}
\partial _{t}T\partial _{r}T=\partial _{t}X\partial _{r}X.  \tag{7}
\end{equation}%
By substituting (5) and (6) in the square of (7), we obtain the mentioned
conformal factor in terms of $T$:%
\begin{equation}
\Omega ^{-1}=f^{-1}\left( \partial _{t}T\right) ^{2}-f\left( \partial
_{r}T\right) ^{2}.  \tag{8}
\end{equation}%
Note that equivalently one can have a similar relation in terms of $X$,
namely $\Omega ^{-1}=f\left( \partial _{r}X\right) ^{2}-f^{-1}\left(
\partial _{t}X\right) ^{2}$. In turn, we use (8) to eliminate $\Omega $ in
(5) and (6), obtaining%
\begin{equation}
\partial _{t}X=f\partial _{r}T  \tag{9}
\end{equation}%
and%
\begin{equation}
\partial _{t}T=f\partial _{r}X,  \tag{10}
\end{equation}%
respectively. By assuming that $T(t,r)=\zeta (t)\psi (r)$ and $%
X(t,r)=\vartheta (t)\chi (r)$, separation of variables imply that%
\begin{equation}
\zeta ^{-1}\frac{d\vartheta }{dt}=\chi ^{-1}f\frac{d\psi }{dr}=a  \tag{11}
\end{equation}%
and%
\begin{equation}
\vartheta ^{-1}\frac{d\zeta }{dt}=\psi ^{-1}f\frac{d\chi }{dr}=b,  \tag{12}
\end{equation}%
with $a$ and $b$ constants. By dividing these relations, one learns that $%
b\vartheta d\vartheta =a\zeta d\zeta $ and $b\psi d\psi =a\chi d\chi $, in
such a way that the new variables are related by%
\begin{equation}
b\vartheta ^{2}-a\zeta ^{2}=A  \tag{13}
\end{equation}%
and%
\begin{equation}
b\psi ^{2}-a\chi ^{2}=B,  \tag{14}
\end{equation}%
where $A$ and $B$ are also constants. As we shall see in next section, this
freedom for choosing the constants admit a plenty of possibilities. Also,
since $d\zeta /dt=b\vartheta $ from (12), differentiating respect to $t$,
together with (11), gives the useful relation%
\begin{equation}
\frac{d^{2}\zeta }{dt^{2}}=ab\zeta .  \tag{15}
\end{equation}%
Notice that in the same manner, $d^{2}\vartheta /dt^{2}=ab\vartheta $ holds.
This allows to englobe three main possibilities: $(ab)$ zero, positive or
negative.

\newpage

\noindent \textbf{3. Regge-Wheeler or tortoise coordinates.}

\smallskip \ 

\noindent The simplest case is when $ab=0$. Assuming $a=0$, we have that
(11) implies that $\vartheta $ and $\psi $ are constants. Without loss of
generality, both can be set to one and by (12) we have that $d\zeta =bdt$,
and\ $\zeta =t$ is a general solution, where we set $b=1$ and selected an
adequate origin of time. Also, the auxiliar variable $\chi $ is defined by%
\begin{equation}
\chi =\int \frac{dr}{f}:=r^{\ast },  \tag{16}
\end{equation}%
with $r^{\ast }$ denoting the Regge-Wheeler or \textit{tortoise coordinate},
for short \cite{MTW}. Recalling that $T=\zeta \psi $ and $X=\vartheta \chi $%
, for $a=0$ we have obtained the simple transformation $T=t$ and $X=r^{\ast
} $. Also, (8) permits to calculate the conformal factor as $\Omega =f$, and
for the metric (3), substitution yields the simple version%
\begin{equation}
ds^{2}=f(-dt^{2}+dr^{\ast 2}).  \tag{17}
\end{equation}%
For $ab=0$, the other possibility is to choose $b=0$, $a=1$, then the
results are mimicked, although now $T$ is changed to $r^{\ast }$ and $X$
becomes $t$, but $\Omega $ turns to be $-f$, returning the same causal
structure of (17). This ambiguity in signs shall be resolved also for the
other cases in order to preserve the association of $T$ and $X$ with the
signature that preserves $\Omega $ positive in the reduced metric $\gamma
_{ab}=\Omega \eta _{ab}$.

\textit{Schwarzschild spacetime}. This case corresponds to $f=1-r_{s}/r$ in
the metric (2), where $r_{s}=2GM$ is the Schwarzschild radius \cite%
{Schwarzschild}. From (16), the tortoise coordinate is given by%
\begin{equation}
r^{\ast }=r+r_{s}\ln \left \vert \frac{r}{r_{s}}-1\right \vert .  \tag{18}
\end{equation}

This transformation is well defined for the exterior solution, and the
absolute value is superfluous. That is, for the patch $r>r_{s}$, decreasing
from $r\rightarrow \infty $ till $r_{s}$, $r^{\ast }$ is ralentized respect
to $r$ (hence the alias \textit{tortoise}), with its range for the mentioned
change decreasing from $+\infty $, passing through $r^{\ast }=0$ when $%
r=[1+W(e^{-1})]r_{s}\approx 1.27846$ $r_{s}$, ($W(x)$ is the W-Lambert
function), and $r^{\ast }$ approaching $-\infty $ for $r\rightarrow r_{s}$.
One goes back in the diagram when taking into account the interior solution,
in such a way that for $r<r_{s}$ on can continue the ingoing path with $%
r^{\ast }$ starting from $-\infty $ when $r=r_{s}$ until $r^{\ast }=0$ when $%
r=0$. Notice that for this patch one should consider $1-{r}/{r_{s}}$ in
the argument of the logarithm in (18).

Other property is that, by taking $t=\pm r^{\ast }+const.$, one obtains a
congruence of ongoing and ingoing radial null geodesics, as can be seen
directly from (17). Otherwise, consider that in Schwarzschild coordinates
the Killing vector $K^{\mu }=\delta _{0}^{\mu }$ induces the conservation of 
$U_{0}=g_{00}\frac{dt}{d\lambda }$, with $\lambda $ the affine parameter.
That is $(1-r_{s}/r)dt/d\lambda =e$, a constant, and substitution in (2)
when $d\theta =d\phi =0$, leads to $dr/d\lambda =\pm e$ \cite{Bronnikov}.
These relations imply that%
\begin{equation}
\frac{dr}{dt}=\mp (1-r_{s}/r),  \tag{19}
\end{equation}%
that is equivalent to integrate again (16) with $t=\pm r^{\ast }+const$.

As we see, this first case is important for several reasons, but for our
purposes it shall define the other class of solutions and also is the
boundary that divides the results of section 4 and those of section
5.\bigskip

\textbf{4. Kruskal-Szekeres and extensions.}

\smallskip \ 

\noindent A less trivial solution is when one considers $ab>0$ in (15). It
can be seen that the transformation admits a subclass of solutions in the
form of exponential functions, or hyperbolic sines and cosines. We shall
assume that $a$ and $b$ are both positive since the results are the same for
both being negative. First consider the exponential case, with $\zeta
=\alpha _{1}e^{\pm \sqrt{ab}t}$ and $\vartheta =\alpha _{2}e^{\pm \sqrt{ab}%
t}\,$, in such a way that (11) and (12) in the form $d\vartheta /dt=a\zeta $
and $d\zeta /dt=b\vartheta $, imply the relation $\alpha _{2}=\sqrt{a/b}%
\alpha _{1}$. Also notice that substituting in (13) yields $A=0$. We shall
choose the positive sign in the exponentials, since similar conclusions are
held when considering the negative signs.

For such a case, one should explore what occurs to the radial auxiliary
functions. If one assumes a similar class for (14), with $B=0$, in such a
way that $\chi =\pm \sqrt{b/a}\psi $, and then by (11) we have that 
\begin{equation}
f\frac{d\psi }{dr}=a\chi  \tag{20}
\end{equation}%
is equal to $\pm \sqrt{ab}\psi $, with solution $\psi =De^{\pm \sqrt{ab}%
r^{\ast }}$.

However, these solutions together are not compatible, since then $T=\zeta
\psi =D\alpha _{1}e^{\sqrt{ab}(t\pm r^{\ast })}$ and $X=\vartheta \chi =\pm
D\alpha _{1}e^{\sqrt{ab}(t\pm r^{\ast })}$. This makes that $-dT^{2}+dX^{2}\ 
$goes to zero in (3), which is compensated by $\Omega $ going to $\infty $
due to (8). Then in (14) we consider $B>0$, which allows to take%
\begin{equation}
\chi =\sqrt{\frac{1}{a}\left( b\psi ^{2}-B\right) },  \tag{21}
\end{equation}%
that converts (20) into%
\begin{equation}
\frac{d\psi }{\sqrt{a\left( b\psi ^{2}-B\right) }}=\frac{dr}{f}.  \tag{22}
\end{equation}%
The solution is $\psi =\sqrt{B/b}\cosh (\sqrt{ab}r^{\ast })$ and then $\chi =%
\sqrt{B/a}\sinh (\sqrt{ab}r^{\ast })$. Considering the exponentials for $%
\zeta $ and $\vartheta $, a possible transformation is

\begin{equation}
T=\sqrt{\frac{B}{b}}\alpha _{1}e^{\sqrt{ab}t}\cosh \left( \sqrt{ab}r^{\ast
}\right)  \tag{23}
\end{equation}%
and%
\begin{equation}
X=\sqrt{\frac{B}{b}}\alpha _{1}e^{\sqrt{ab}t}\sinh \left( \sqrt{ab}r^{\ast
}\right) ,  \tag{24}
\end{equation}%
Also, inserting (23) in (8), the conformal factor becomes%
\begin{equation}
\Omega =(Bb\alpha _{1}^{2})^{-1}e^{-2\sqrt{ab}t},  \tag{25}
\end{equation}%
and substitution in (3) clearly agrees with (17). Consider again the
Schwarzschild case. Selecting $\sqrt{B/b}\alpha _{1}=1$ and $\sqrt{ab}%
=r_{s}^{-1}$, we have the defining hyperbolae for constant Schwarzschild
time $t$:%
\begin{equation}
T^{2}-X^{2}=e^{\frac{2t}{r_{s}}},  \tag{26}
\end{equation}%
while for constant $r$ we have straight lines defined by%
\begin{equation}
X=T\tanh \left( \frac{r^{\ast }}{r_{s}}\right) .  \tag{27}
\end{equation}%
The structure of the spacetime is characterized in this coordinates in Fig.
1.

\begin{figure}[H]
    \centering
    \includegraphics[scale=.6]{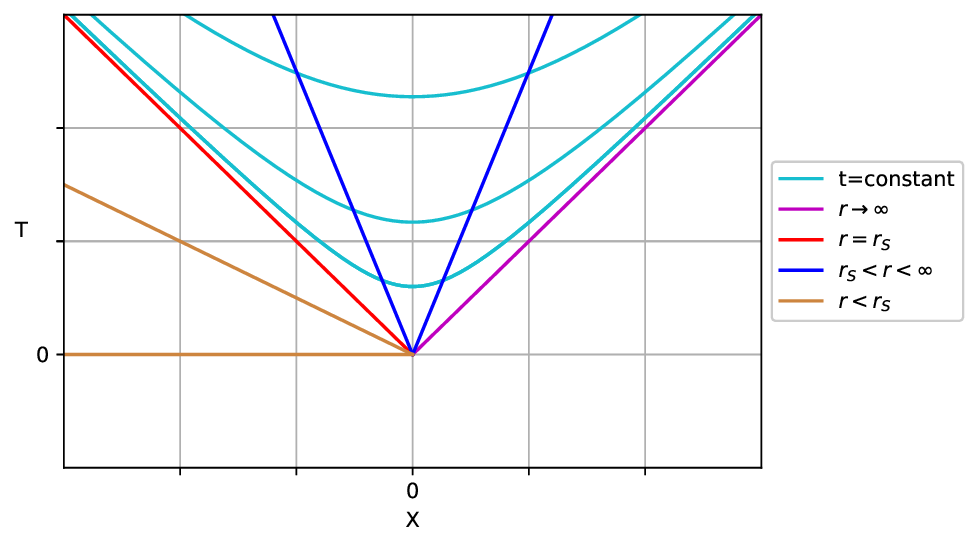}
    \caption{Diagram of hyperbolic coordinates for Schwarzschild spacetime, where lines
    represent constant r and only exterior and interior solution are shown.}
    \label{fig:pseudokruskal}
\end{figure}

For the exterior solution, starting from $r\rightarrow \infty $, at $%
r=const. $ we have\ straight lines that vary from $45%
{{}^\circ}%
$ with angle decreasing continuously and making $X=0$ when $r\approx 1.27846$ 
$r_{s}$, till one reaches the line at $-45%
{{}^\circ}%
$ ( $r=r_{s}$). If one is willing to include the interior solution in the
same diagram, one could associate all the lines with angle $\geq -45%
{{}^\circ}%
$ again with the values of $r<r_{s}$, similar to what occurs for $r^{\ast }$
[see comments below (18)]. Instead, we represent the interior solution by
choosing a sign and interchanging the hyperbolic functions in (23) and (24),
which makes that the lines at constant $r$ vary continuously from the red
line in Fig. 1 till $T=0$, corresponding to $r=0$. We have omitted the
horizontal hyperbolas at constant $t$. Here, the exterior region is the
upper part inside the `cone', from $X=T$ to $X=-T$, and the interior region
is delimited in the second quadrant by the lines $X=-T$ and $T=0$. The other
two regions of a maximal extension, not shown, are merely a reflection.

Getting back to the distinct possibilities for $ab>0$, we have that $\zeta $
and $\vartheta \ $also admits hyperbolic sines and cosines as solutions. An
appropriate selection is $\zeta =\alpha _{1}\sinh \left( \sqrt{ab}t\right) $
and $\vartheta =\alpha _{2}\cosh \left( \sqrt{ab}t\right) $. Now, (13) and
the other relations are satisfied with $\alpha _{2}=\sqrt{a/b}\alpha _{1}=%
\sqrt{A/b}$, that is for $A$ positive. Although hyperbolic functions of $%
r^{\ast }$ are allowed, the simplest case is choosing the exponentials
mentioned below Eq. (20), namely $\psi =De^{\sqrt{ab}r^{\ast }}$ and $\chi =%
\sqrt{b/a}\psi $ -a positive sign was chosen. Inserting in $T$ and $X$, we
have,%
\begin{equation}
T=\alpha _{1}De^{\sqrt{ab}r^{\ast }}\sinh \left( \sqrt{ab}t\right)  \tag{28}
\end{equation}%
and%
\begin{equation}
X=\alpha _{1}De^{\sqrt{ab}r^{\ast }}\cosh \left( \sqrt{ab}t\right) , 
\tag{29}
\end{equation}%
respectively. As before, $\alpha _{1}D$ can be set equal to one without loss
of generality, and for Schwarzschild solution it turns out to be useful $%
\sqrt{ab}=\left( 2r_{s}\right) ^{-1}$, since then from (18) one obtains $e^{%
\sqrt{ab}r^{\ast }}=e^{r/(2r_{s})}\sqrt{r/r_{s}-1}$ for the exterior
solution, and where the defining surfaces now are hyperbolas at constant $r$%
, with%
\begin{equation}
X^{2}-T^{2}=e^{\frac{r}{r_{s}}}\left( r/r_{s}-1\right) ,  \tag{30}
\end{equation}%
That is, by choosing adequately the constant $ab$, we have obtained the usual
Kruskal solution \cite{Kruskal}\cite{Szekeres}. Moreover, from the
perspective of the present work, (30) is the minimal solution that uniquely
represents the interior and exterior solution to the event horizon \cite{MTW}%
. The interior is included in the other transformations by selecting $e^{%
\sqrt{ab}r^{\ast }}=e^{r^{\ast }/(2r_{s})}\sqrt{1-r/r_{s}}$ due to (18) as
well as by interchanging the hyperbolic functions in (28) and (29). See
Figure 2, where the maximal extension is shown, and lines at $t=const.$ are
omitted.

\begin{figure}[H]
    \centering
    \includegraphics[scale=.5]{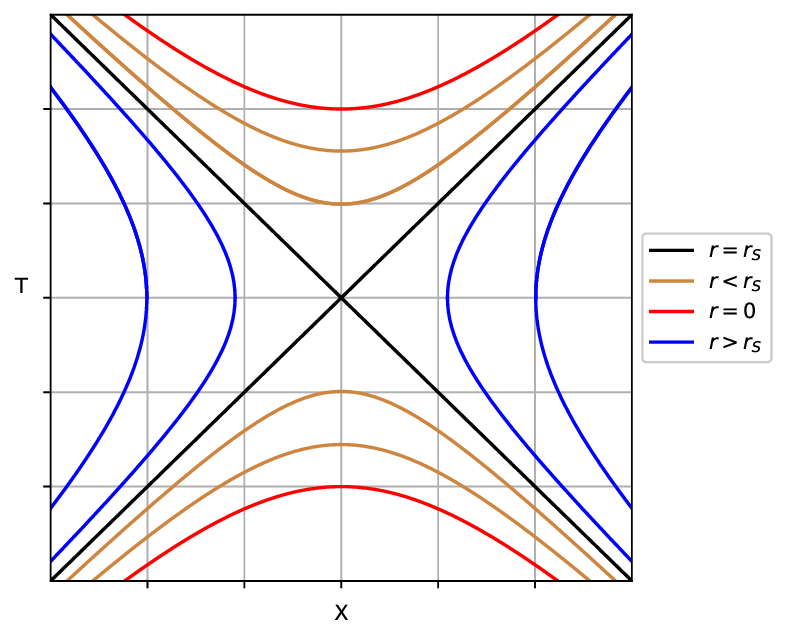}
    \caption{Kruskal diagram for the Schwarzschild case. The selection of signs
    in the constants allows to include the exterior and interior solution in
    the same diagram, as well as to extend it as customary.}
    \label{fig:kruskal}
\end{figure}

Recall the discussion from Section 3: negative $r^{\ast }$ can be describing
interior or exterior solutions, and the correspondence is not one to one.
However, both in this Kruskal-Szekeres case with hyperbolas at constant $r$,
as well as the Kruskal-type variation of Fig. 1 with lines in place of
hyperbolas, we can select the signs in the integration constants, in such a
way that exterior and interior solutions are uniquely represented in the
same diagram.

As we shall see, the final case $(ab<0)$ induces coordinates that don't
share this property, but will be interesting in other respects.\bigskip

\textbf{5. Pulsating coordinates.}

\smallskip \ 

\noindent Now we turn \ our attention for the final possibility dictated by
the relations at the end of section 2. For $ab<0$ in (15), we have that it
allows to choose%
\begin{equation}
\zeta =\beta _{1}\sin (\omega t),  \tag{31}
\end{equation}%
where we have defined $\omega =\sqrt{-ab}$. We shall focus just in the case
of having $a<0$, $b>0$, since $b<0$ would just interchange the role of the $%
X $ and $T$ coordinates and the sign of the conformal factor. Then, relation
(12) in the form $\vartheta =b^{-1}d\zeta /dt$ leads in this case to%
\begin{equation}
\vartheta =\beta _{2}\cos (\omega t),  \tag{32}
\end{equation}%
where $\beta _{2}=\sqrt{-a/b}\beta _{1}$. This in turn implies that the
constant $A$ in (13) is equal to $-a\beta _{1}^{2}$, that is positive. One
can readily see that in (14) we also have $B>0$, and then we can select $%
\sqrt{b}\psi =\sqrt{B+a\chi ^{2}}$. Inserting this in (12) stated as $fd\chi
/dr=b\psi $, we have to integrate the relation%
\begin{equation}
\frac{d\chi }{\sqrt{b(B+a\chi ^{2})}}=\frac{dr}{f}.  \tag{33}
\end{equation}%
The result is%
\begin{equation}
\chi =\sqrt{\frac{B}{-a}}\sin (\omega r^{\ast }+\varphi _{0})  \tag{34}
\end{equation}%
and consequently%
\begin{equation}
\psi =\sqrt{\frac{B}{b}}\cos (\omega r^{\ast }+\varphi _{0}).  \tag{35}
\end{equation}%
Since we could have chosen the negative sign for $\psi $, and $\varphi _{0}$
is arbitrary, this gives us the possibility of choosing sines or cosines for 
$\chi $ and $\psi $. For instance, by choosing $\varphi _{0}=90%
{{}^\circ}%
$ and a sign in $\psi $, previous relations yield the coordinate
transformations $T=\frac{\sqrt{AB}}{\omega }\sin (\omega t)\sin (\omega
r^{\ast })$ and $X=\frac{\sqrt{AB}}{\omega }\cos (\omega t)\cos (\omega
r^{\ast })$. By taking the differential and by substituting in (8), one can
see that (17) is satisfied when $\Omega =2f/(AB)\left[ \cos (2\omega t)-\cos
(2\omega r^{\ast })\right] ^{-1}$. However, here $\Omega $ goes to infinity
whenever $r^{\ast }=t$. Instead, we choose $\varphi _{0}=0$ and negative
sign in $\psi $, which yields a similar selection:%
\begin{equation}
T=\frac{\sqrt{AB}}{\omega }\sin (\omega t)\cos (\omega r^{\ast })  \tag{36}
\end{equation}%
and%
\begin{equation}
X=\frac{\sqrt{AB}}{\omega }\cos (\omega t)\sin (\omega r^{\ast }).  \tag{37}
\end{equation}%
Also, substitution in (17) yields the conformal factor%
\begin{equation}
\Omega =\frac{2f}{AB\left[ \cos (2\omega t)+\cos (2\omega r^{\ast })\right] }%
.  \tag{38}
\end{equation}%
By comparing with the Kruskal-Szekeres or related coordinates such as those
of Section 4, several differences arise. In place of having hyperbolas and
lines at constant $r$ or $t$, here in general one has ellipses or
circumferences, since the curves at constant $r$ are given by%
\begin{equation}
\frac{X^{2}}{\sin ^{2}(\omega r^{\ast })}+\frac{T^{2}}{\cos ^{2}(\omega
r^{\ast })}=\frac{AB}{\omega ^{2}},  \tag{39}
\end{equation}%
while at $t$ constant we have%
\begin{equation}
\frac{X^{2}}{\cos ^{2}(\omega t)}+\frac{T^{2}}{\sin ^{2}(\omega t)}=\frac{AB%
}{\omega ^{2}}.  \tag{40}
\end{equation}%
Fig. (3) shows the mentioned behavior considering that the curves
corresponds to constant $r$ (we chose $AB=1$ for simplicity).

\begin{figure}[H]
    \centering
    \includegraphics[scale=.6]{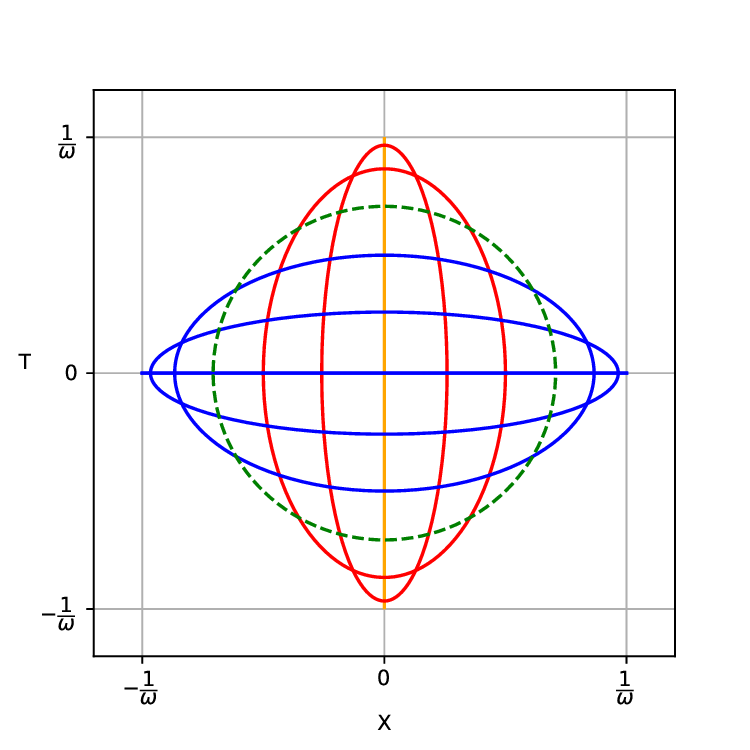}
    \caption{Pulsating coordinates, with oscillatory character of the coordinates and
    compactification in representation at both r and t constants.}
    \label{fig:pulsating}
\end{figure}

Here, one has that all the spacetime is compactified to maximum limits of $%
\pm 1$ in the coordinates $X$ and $T$. Notice that the presence of any
horizon makes $r^{\ast }$ to vary in an infinite range in any casual patch
of the spacetime described by (2). This in turn implies that the surfaces
corresponding to constant $r$ in pulsating coordinates make infinite
oscillations between vertical and horizontal ellipses, passing through
vertical lines, circumferences or horizontal lines for the special values $%
\omega r^{\ast }=n\pi $, $\omega r^{\ast }=(4n+1)\pi /4$ or $\omega r^{\ast
}=(2n+1)\pi /2$, with $n$ a non-negative integer.

Clearly, this series of oscillations can be avoided for any finite domain of 
$r$ that does not include the event horizon. For instance, in the
Schwarzschild case, take the decreasing values between the photon sphere ($%
r=1.5r_{s}$) and the zero of the tortoise coordinate. By choosing $\omega
=r_{s}^{-1}$, then we have first an horizontal ellipse with semimajor axis $%
X=0.72211$ that converts into a circle when $r\approx 1.492882544$ $r_{s}$
and then into vertical ellipses that asymptotically tend to a vertical line
when $r\rightarrow 1.27846$ $r_{s}$.

Other case of interest include metrics that describe cosmological models.
For instance, take the FLRW models%
\begin{equation}
ds^{2}=-d\tau ^{2}+a^{2}\left[ \frac{d\rho ^{2}}{1-k\rho ^{2}}+\rho
^{2}\left( d\theta ^{2}+\sin ^{2}\theta d\phi ^{2}\right) \right] ,  \tag{41}
\end{equation}%
where $\tau $ is the cosmological time measured by an observer with comoving
radial coordinate $\rho $ \cite{Peebles}. The subclass of FLRW solutions
that can be converted to (2) are only those for which $f=1-\Gamma r^{2}$ and
where the only possible contribution to the energy momentum tensor is vacuum
energy \cite{Edgar}\cite{Florides}. For the distinct possibilities allowed
for the cosmological constant $\Lambda =3\Gamma $ and the curvature
parameter $k$, this subclass of solutions includes de-Sitter or
Anti-de-Sitter spaces, Lanczsos universe, Milne model or even Minkowski
space \cite{Melia}\cite{Mitra}. Take for instance the de-Sitter solution,
where $\Gamma >0$ and $k=0$, with scale parameter $a=\exp (\sqrt{\Gamma }%
\tau )$. Here, the solution of (16) for the tortoise coordinate is $r^{\ast
}=\Gamma ^{-1/2}\tanh ^{-1}(\Gamma ^{1/2}r)$, that induces again infinite
oscillations in the variable $X~$(at $t$ constant), see Fig. 4.

\begin{figure}[H]
    \centering
    \includegraphics[scale=.5]{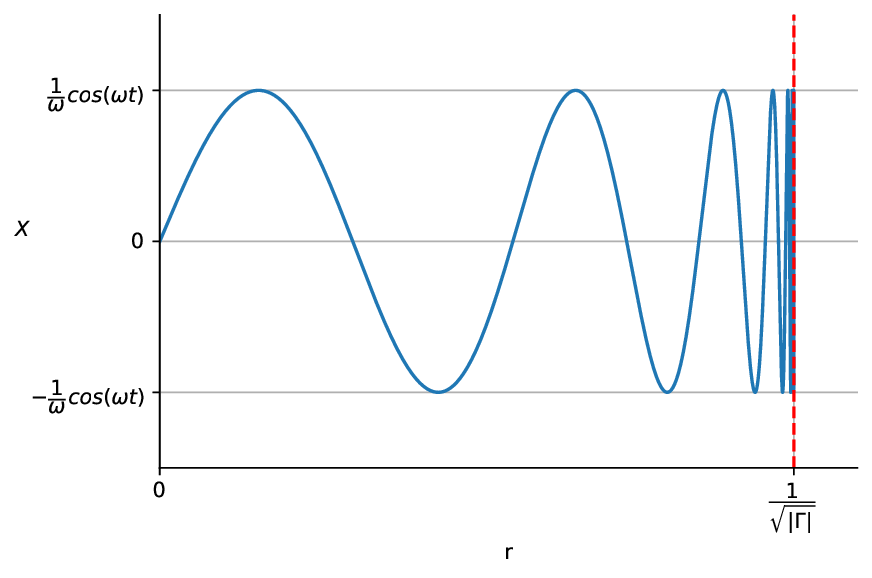}
    \caption{\small{Oscillations in the pulsating coordinate X, that are more pronunciated near a causal
    horizon (in this case cosmological horizon).}}
    \label{fig:Xelipse}
\end{figure}

These oscillations become more dramatic as $r\rightarrow \Gamma ^{-1/2}$,
the cosmological horizon. This in turn induces the mentioned oscillations
between ellipses in the $(X,T)$ diagram.

Now consider Anti-de-Sitter space, where both $\Gamma $ and $k$ are
negative. The scale parameter is obtained from the first Friedmann equation
as $a(\tau )=1/\sqrt{\left \vert \Gamma \right \vert }\sin (\sqrt{%
\left
\vert \Gamma \right \vert }\tau )$ \cite{Edgar}. Also, solving for
the tortoise coordinate (16) we obtain $r^{\ast }=1/\sqrt{\left \vert \Gamma
\right \vert }\tan ^{-1}(\sqrt{\left \vert \Gamma \right \vert }r)$, and
then for this case $r^{\ast }$ goes from $0$ to $\pi /\left( 2\sqrt{%
\left
\vert \Gamma \right \vert }\right) $ when $r$ goes from $0$ to $%
\infty $. By choosing $\omega =\sqrt{\left \vert \Gamma \right \vert }$,
then (39) leads to a vertical line for $r=0$ that converts into vertical
ellipses in $0<r<1/\sqrt{\left \vert \Gamma \right \vert }$ in Fig. 3
(yellow lines), then into a circumference of radius $1/\sqrt{2}$and from
there vertical ellipses for $r>1/\sqrt{\left \vert \Gamma \right \vert }$
that approach the horizontal line $T=0$ as $r\rightarrow \infty $, which in
Fig. 3 appear in blue lines.

Similar argument occurs to the surfaces at $t$ constant, where in agreement
with (40) the change is from horizontal ellipses to vertical ones.\bigskip

\noindent \textbf{6. Final remarks.}

\smallskip \ 

\noindent In this work, we have studied an approach that relates several
types of transformations for spherically symmetric spaces, that can be put in
the form%
\begin{equation}
dS^{2}=-fdt^{2}+f^{-1}dr^{2}+r^{2}d\theta ^{2}+r^{2}\sin ^{2}\theta d\phi
^{2}.  \tag{42}
\end{equation}%
This includes spaces associated describing black holes such as Schwarzschild,
Reissner-Norstr\"{o}m, but also allows for cosmological solutions such as
de-Sitter, Anti-de-Sitter or Schwarzschild-de-Sitter, among others. Our
approach relates several possible transformations to a flat conformal flat
(1+1)-type when considering constant angles. The developments of Section 2,
and in particular the sign of the product $ab$ in Eq. (15), led us to derive
three coordinate solutions: tortoise, Kruskal-Szekeres type and pulsating
coordinates.

In Section 3 we obtained the tortoise case as the unique possibility for $%
ab=0$, and also mentioned some properties of the tortoise coordinates that
where relevant for the other two general cases. Section 4 was devoted to
analyze the case $ab>0$, where several combinations of exponential and
hyperbolic functions appear. We mentioned two selections, one where lines $%
X\propto T$ denote spheres with constant $r$, as well as the usual
Kruskal-Szekeres transformation. Both elections allow a distinction of the
interior and exterior regions in the same diagram.

As a final part, in Section 5 we introduced the notion of pulsating
coordinates as an unexplored possibility for $ab<0$, which leads also to
several possibilities for defining $X$ and $T$ in terms of trigonometric
functions, from which we have selected an appropriate one, described by the
relations (36)-(40). We argue that these solutions have the advantage of
compactifying the space-time in a different way of the Penrose-Carter
compactification. For instance, contrast the exponential grow implied by
(28) and (29) in the the Kruskal diagram with the oscillatory behavior in
the pulsating coordinates. This new system of coordinates present infinite
oscillations between hyperbolas when taking into account all the domain of $%
r $ in a causally connected patch with an event horizon.

Throughout the article we have mentioned the Schwarzchild spacetime as the
prototype for the analysis in all the coordinates presented. However, for
the case of pulsating coordinates we have also contrasted what occurs in two
cosmological models of interest, namely de-Sitter (dS) and Anti-de-Sitter
(AdS). We find that the behavior is better for the AdS solution, since in
this case one can obtain a unique transition from vertical to horizontal
ellipses. Given the fact that the crossing of ellipses with the $X$-axis is
unique for any value of $r$ going from $X=0$ when $r=0$ to $X\rightarrow 1$
when $r\rightarrow \infty $ ($r^{\ast }\rightarrow 1/\sqrt{\left \vert
\Gamma \right \vert }$), we have a natural dimensional reduction for the AdS
spacetime that emulates the Poincar\'{e} mapping, useful in the study of
dynamical systems \cite{Smale}. This line of thinking, as well as the
possibility of relating exterior and interior solutions in black holes and
cosmology \cite{Nieto}, or the specific use for other relevant metrics (see 
\cite{Graves}-\cite{Topo} and References therein), are left as possible
future works.

\bigskip \ 

\noindent \textbf{Acknowledgments: }ASR and BMO acknowledge a graduate
fellowship grant by CONACYT-Mexico.

\newpage

\noindent \textbf{Appendix. Existence of conformal flat transformation in
(1+1) dimensions.}

\smallskip \ 

\noindent Any (1+1) metric $g_{ab}dx^{a}dx^{b}$ can be put in a conformal
flat way $\Omega \eta _{ab}dx^{a\prime }dx^{b\prime }$, where $\eta
_{ab}=diag(-1,1)$. The plainer way to see it is to consider null coordinates 
$v$ and $u$. Here $\mathbf{g}(\partial _{v},\partial _{v})=0$, where $%
\partial _{v}$ is the basis for any null vector in the v-direction. Since
the dual version in terms of the inverse metric is $\mathbf{g}^{-1}(dv,dv)=0$%
, and given the one-form $dv=(\partial v/\partial x^{a})dx^{a}$, this is the
same as $g^{ab}\partial v/\partial x^{a}\partial v/\partial x^{b}=0$, that
can be seen as the definition of a null coordinate, and analog definitions
for the other null coordinate $u$ \cite{Inverno}. Defining $x^{0^{\prime
}}=v $ and $x^{1^{\prime }}=u$, those relations can be casted as the
transformation of the metric to new coordinates, in the form%
\begin{equation}
g^{\alpha \beta }\frac{\partial v}{\partial x^{\alpha }}\frac{\partial v}{%
\partial x^{\beta }}=g^{0^{\prime }0^{\prime }}=0  \tag{A.1}
\end{equation}%
and%
\begin{equation}
g^{\alpha \beta }\frac{\partial u}{\partial x^{\alpha }}\frac{\partial u}{%
\partial x^{\beta }}=g^{1^{\prime }1^{\prime }}=0.  \tag{A.2}
\end{equation}

Now, if $g^{ab}$ has diagonal elements equal to zero, so does $g_{ab}$,
since in this case $g^{00}=(g^{01})^{2}g_{11}$ and $g^{01}\neq 0$ for the
inverse to exist. It follows that $g_{11}=0$ and similar argument yields $%
g_{00}=0$. Then, in terms of null coordinates the system simplifies to 
\begin{equation}
ds^{2}=g_{\alpha ^{\prime }\beta ^{\prime }}dx^{\alpha ^{\prime }}dx^{\beta
^{\prime }}=g_{0^{\prime }1^{\prime }}(dvdu+dudv).  \tag{A.3}
\end{equation}%
Rotating coordinates by means of $T=v+u$ and $X=v-u$, the metric takes the
assumed conformal form%
\begin{equation}
ds^{2}=\Omega (-dT^{2}+dX^{2}),  \tag{A.4}
\end{equation}%
where $\Omega =-2g_{0^{\prime }1^{\prime }}$.

\end{document}